\pdfoutput=1

\documentclass[peerreview]{IEEEtran}
\usepackage{cite}

\ifCLASSINFOpdf
\else
\fi
\usepackage{amsmath}
\usepackage{bm}
\usepackage{amsfonts}
\usepackage{array}
\usepackage{mathtools}
\begin{document}

\title{Adaptivity provably helps: information-theoretic limits on $l_0$ cost of non-adaptive sensing}
%
\author{\Large{Sanghamitra~Dutta and Pulkit~Grover}
\thanks{Sanghamitra Dutta and Pulkit Grover are with Department of Electrical and Computer Engineering, Carnegie Mellon University, Pittsburgh, PA 15213, USA. Emails:- {sanghamd, pgrover}@andrew.cmu.edu.}
\thanks{This paper has been accepted at the IEEE International Symposium on Information Theory, 2016.}}

\maketitle
\begin{abstract}
 The advantages of adaptivity and feedback are of immense interest in signal processing and communication with many positive and negative results. Although it is established that adaptivity does not offer substantial reductions in minimax mean square error for a fixed number of measurements, existing results have shown several advantages of adaptivity in complexity of reconstruction, accuracy of support detection, and gain in signal-to-noise ratio, under constraints on sensing energy. Sensing energy has often been measured in terms of the Frobenius Norm of the sensing matrix. This paper uses a different metric that we call the $l_0$ cost of a sensing matrix-- to quantify the complexity of sensing. Thus sparse sensing matrices have a lower cost. We derive information-theoretic lower bounds on the $l_0$ cost that hold for any non-adaptive sensing strategy. We establish that any non-adaptive sensing strategy must incur an $l_0$ cost of a $\Theta\left( N \log_2(N)\right) $ to reconstruct an $N$-dimensional, one--sparse signal when the number of measurements are limited to $\Theta\left(\log_2 (N)\right)$. In comparison, bisection-type adaptive strategies only require an $l_0$ cost of at most $\mathcal{O}(N)$ for equal order of measurements. The problem has an interesting interpretation as a sphere packing problem in a multidimensional space, such that all the sphere centres have minimum non-zero co-ordinates. We also discuss the variation in $l_0$ cost as the number of measurements increase from $\Theta\left(\log_2 (N)\right)$ to $\Theta\left(N\right)$.
\end{abstract}


%


\section{Introduction}


\IEEEPARstart{T}{he} problem of compressed sensing and sparse reconstruction \cite{candes2008introduction} \cite{donoho2006compressed} has drawn tremendous interest over the past decade due to its potential applications in sub-Nyquist sampling, imaging, biomedical signal processing, astronomy and geophysics. Consider the problem of reconstructing an $N$ dimensional sparse signal vector $\bm{x}$ from its noisy $M$ dimensional compressed measurement vector $\bm{y}$:
\begin{equation}\label{eq1}
\bm{y}=\bm{A}\bm{x} + \bm{z}
\end{equation}
Here $\bm{A}$ is the $M \times N$ dimensional sensing matrix with $ M \ll N$ and $\bm{z}$ is the noise vector. We assume $\bm{z} \sim \mathcal{N}(0,\bm{I}_{M \times M}) $. The signal $\bm{x}$ is said to be $K$--sparse (Usually $K \ll N$) if $ ||\bm{x}||_0 \leq K$. A detailed review of various sparse reconstruction algorithms proposed over the past decade can be found in \cite{tropp2010computational}.

It appears intuitive that during the acquisition of compressive measurements of an unknown signal $\bm{x}$, ``choosing an adaptive or sequential strategy that cleverly selects the next rows of sensing matrix based on what has been previously observed"\cite{arias2013fundamental} might help in easily reconstructing the unknown signal as compared to a non-adaptive strategy. In a non-adaptive strategy, the entire measurement vector $\bm{y}$ is acquired in one-shot and the rows of sensing matrix $\bm{A}$ are not chosen based on any prior or acquired information. In this paper, we choose the $l_0$ cost (similar to ``$l_0$ norm"), \textit{i.e.}, number of non-zero entries of the sensing matrix as our measure of sensing complexity for comparison between adaptive \cite{malloy2014near} \cite{davenport2012compressive} and non-adaptive strategies. 

In this paper, we derive a general information-theoretic lower bound on the $l_0$ cost of reconstruction of a one--sparse signal that hold for any non-adaptive sensing matrix and demonstrate that adaptivity can achieve a lower $l_0$ cost as compared to any non-adaptive strategy, by a factor that diverges to infinity for large $N$. It is well established that both adaptive \cite{malloy2014near} and non-adaptive \cite{baraniuk2008simple} \cite{do2010lower} sensing strategies require at least $\Theta\left(\log_2(N)\right)$ measurements \cite{grover2012fundamental} \cite{aksoylar2014information}for the reconstruction of a one--sparse signal. We show that in the reconstruction of a one--sparse signal, any adaptive strategy would require an $l_0$ cost of $\Theta\left({N\log_2(N)}\right) $ as compared to bisection-type\cite{malloy2014near} adaptive strategies that require at most an $\mathcal{O}(N) $ $l_0$ cost when the number of measurements are limited to $\Theta({\log_2(N)})$. Our problem translates into a novel sphere packing problem within an $M$ dimensional sphere with an additional constraint of minimizing the $l_0$ cost of the sphere centres. We also discuss scenarios when the number of measurements varies from $\Theta(\log_2(N))$ to $\Theta(\log_2(N))$. Lastly, we provide outlines on extension of our results to $K$--sparse signals.

A sparser sensing matrix would offer significant advantages in terms of implementation, storage and reconstruction as discussed in \cite{gilbert2010sparse}. However, this line of work is mostly focused on the design of non-adaptive, sparse matrices and efficient reconstruction algorithms, rather than the derivation of fundamental limits. Adding on to the benefits of sparse matrices in \cite{gilbert2010sparse}, one can view the process of acquiring a measurement, \textit{i.e.}, computing the dot product of the signal with a row of the sensing matrix as a ``filtering-type" operation. The magnitude of the entries in the sensing matrix then corresponds to amplification/scaling during sample acquisition, while a zero value implies that no filter-tap (multiplier) is turned on. Then the $l_0$ cost would actually correspond to the number of multipliers required during compressive acquisition and thus have a direct implication on the sensing cost. A higher number of multipliers would also add more noise to the system and also increase the acquisition complexity. Similarly, in group testing/pooling type applications, the $l_0$ cost would correspond to the number of mixing operations required. 

While not in the context of sparsity of sensing matrices, there is a body of literature that addresses the question of whether and when adaptivity is really helpful. Arias-Castro, Candes and Davenport \cite{arias2013fundamental} derive a minimax (worst case) lower bound on the mean square error for any sensing strategy (adaptive/non-adaptive). While they do not show any explicit advantage of adaptivity as in this paper, the authors argue that the potential benefits in Mean Square Error offered by adaptivity ``in the worst case" are within $\mathcal{O}(\log(N))$ over non-adaptive, random strategies. In this paper, we show an explicit advantage to adaptivity as compared to \textit{any} non-adaptive strategy and also highlight the costs of using random matrices in terms of $l_0$ cost. On the other hand, Malloy and Nowak\cite{malloy2014near} point out various advantages to adaptivity that include reducing reconstruction complexity (as does feedback in noisy channels) and also in providing gain in signal-to-noise ratio under limitations of sensing energy. They show that ``bisection-type" adaptive strategies can recover the support of the unknown $K$--sparse signal with minimum amplitude  $\Omega\left(\sqrt{\frac{N}{B}\log(K)}\right)$ as compared to traditional non-adaptive strategies that require an amplitude of $\Omega\left(\sqrt{\frac{N}{B}\log(N)}\right)$\cite{wang2010information} \cite{aeron2010information} for the same number of measurements, \textit{i.e.}, $\Theta(K\log_2(N))$. Here $\sqrt{B}$ is a constraint on the Frobenius norm of the sensing matrix. In a different scenario where $M>N$, Haupt, Castro and Nowak \cite{haupt2011distilled} establish that adaptivity in sensing can recover signals with "vanishingly small" amplitudes as compared to random non-adaptive strategies, under sensing energy constraints. 
Some other related references that show advantages of adaptivity in specific scenarios can be found in \cite{malloy2011limits}, \cite{castro2008finding} and \cite{braun2014info}.

\begin{figure}[!t]
\centering
\includegraphics[height=3 in]{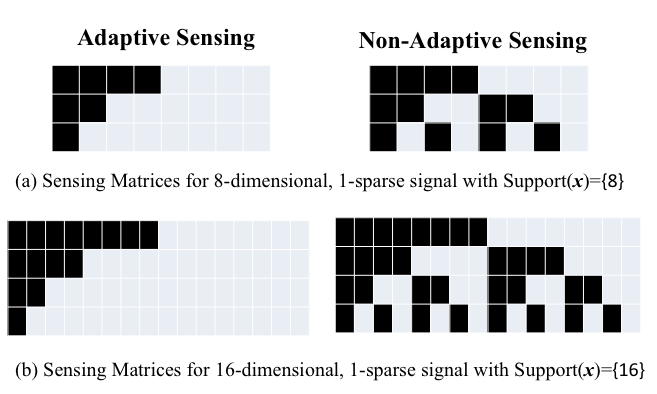}

\caption{The figure provides intuition on why adaptive sensing could have fundamentally sparser matrices than nonadaptive sensing. The matrices on the left use an adaptive bisection strategy to narrow down the search space in $\Theta (\log_2 N)$ steps, requiring $\Theta(\log_2 N)$ measurements. Interestingly, the construction on the right can simulate the bisection strategy nonadaptively. However, the cost is increased density of the matrices.}
\label{fig1}
\end{figure}

\textit{Why would one think adaptivity can make the sensing matrix sparser?} Fig.\ref{fig1} illustrates a scenario of ``bisection-type" adaptive sensing in comparison to non-adaptive sensing as the sensing matrices get sparser. It provides an intuition that while the number of measurements remain the same (at least in order sense), even with the sparsest non-adaptive sensing matrices, adaptivity always incurs a lower $l_0$ cost than non-adaptive sensing. References in favour of adaptivity \cite{malloy2014near} have cited existing information-theoretic bounds\cite{aeron2010information}\cite{wang2010information} on SNR and number of measurements for non-adaptive compressive sensing that are mostly based on certain deterministic matrices or random, Gaussian matrices. However, Gaussian matrices are dense with an $l_0$ cost of $N M$. Adaptivity can allocate non-zero entries in the sensing matrix selectively, like a binary search. The intuition is most clear for one--sparse signals, and hence those are discussed first.

\section{Notation and Problem Formulation}


Let the $l_0$ cost of a sensing matrix $\bm{A}=[A_1 A_2 \ldots A_N]$ in problem (\ref{eq1}) be defined as:
\begin{equation}
||\bm{A}||_{0,0} = \sum_{i=1}^N ||A_i||_0 = \sum_{i=1}^N \sum_{j=1}^M \mathbb{I}(A(j,i) \neq 0)
\end{equation} 
Here $A_i$ denotes the $i$-th column of $\bm{A}$ and $\mathbb{I}(.)$ denotes the indicator function. 
We say a function $F(N)=\Omega(g(N))$ if there exist a positive constant $C_0$ such that, $F(N)\geq C_0 g(N)$ as $N \to \infty$. It implies that asymptotically, $g(N)$ is not dominated by $F(N)$. Similarly, $F(N)=\Theta(g(N))$ if there exist positive constants $C_1,C_2$ such that, $C_1 g(N) \leq F(N) \leq C_2 g(N)$. It implies that neither $F(N)$ nor $g(N)$ are asymptotically dominated by each other. Lastly, $F(N)=\mathcal{O}(g(N))$ if $F(N) \leq C_3 g(N)$ for some positive constant $C_3$.

We derive lower bounds on $||\bm{A}||_{0,0}$ for non-adaptive sensing that asymptotically dominate upper bounds for adaptive sensing. In Theorem 1 we consider the case of reconstruction of a one sparse signal $\bm{x}$ using a \textit{binary} sensing matrix $\bm{A}$ in the absence of noise. In Theorem 2, we extend our analysis to real-valued sensing matrices for reconstructing a  one sparse signal $\bm{x}$ with a target probability of error in the presence of additive  white Gaussian noise. Note that we have assumed $M$ to vary as a logarithmic function of $N$, when $N$ is large. 




\section{Main Results}


Our first theorem establishes an information-theoretic lower bound on the $l_0$ cost of non-adaptive sensing for \textit{binary} sensing matrices in the absence of noise when the number of measurements are limited to $\Theta\left( \log_2 (N)\right)$.

\newtheorem{theorem}{Theorem}
\begin{theorem} 
 For the reconstruction of an arbitrary one--sparse signal with amplitude $\mu$ using \textit{binary} sensing matrices with number of measurements $M =T \log_2 (N)$ where $T$ is a constant greater or equal to $1$, any non-adaptive sensing strategy incurs an $l_0$ cost of $ \Theta \left(N\log_2(N)\right) $.
\end{theorem}

\textit{Proof Sketch}:
Consider the recovery of the support of a one--sparse vector $\bm{x}$, from $\bm{y}$ and $\bm{A}$ when $ M=\Theta \left(\log_2(N)\right) $. Then, any non-adaptive strategy would at least require the columns $\{A_i\}$ of $\bm{A}$ to be distinct. Notice $$\bm{A x}=[A_1,\cdot \ \cdot, A_N]\begin{bmatrix}
x_1\\
\cdot\\
\cdot\\
x_N
\end{bmatrix}=\sum_{i=1}^N A_ix_i$$
If two columns (say $i_1$ and $i_2$) of $\bm{A}$ are equal, then two one--sparse $\bm{x}$ vectors with respective supports $\{i_1\}$ and $\{i_2\}$ become indistinguishable. 
Among all \textit{binary} sensing matrices $\bm{A}$ with $N$ unique columns, we consider one with minimum $l_0$ cost and derive lower bounds on its $l_0$ cost. Such a \textit{binary} sensing matrix $\bm{A}$ with minimum $l_0$ cost can be easily designed by choosing the columns in ascending order of $l_0$ cost, \textit{i.e.}, first all possible columns with $l_0$ cost $1$, then all possible columns with $l_0$ cost $2$ and so on until we get $N$ unique columns. We show that $||\bm{A}||_{0,0}$ is always $\Theta(N\log_2(N))$. 

\emph{Note that for $M < \log_2(N)$, it is not possible to construct binary sensing matrices with unique non-zero columns. So we assume $M =T \log_2 (N)$ where $T \geq 1$.} 
Our next theorem establishes a lower bound on the $l_0$ cost of any non-adaptive sensing matrix in the presence of noise, when $||A_i||_2$ is bounded. For non-adaptive strategies, limitations of sensing energy \cite{malloy2014near} usually enforce upper bounds on $||A_i||_2$, thus justifying the assumptions of \textit{Theorem 2}.

\begin{theorem}
For the reconstruction of an arbitrary one--sparse signal with amplitude $\mu$ in the presence of Gaussian Noise of unit variance, using real-valued sensing matrices, under the constraints that $M =T \log_2 (N)$ where $T$ is a constant greater or equal to $1$ and $||A_i||_2$ is bounded by  $\tau$, any non-adaptive sensing strategy cannot achieve a target probability of error $\epsilon$  without incurring an $l_0$ cost of  $\Theta\left(N \log_2(N)\right)$, provided that the allowed bound on $||A_i||_2$ \textit{i.e.}, $\tau=\Theta(Q^{-1}(\epsilon)/\mu) $ .
\end{theorem}

\textit{Proof Sketch:}
First consider the case of one--sparse signal $\bm{x}$ with non-zero value equal to $\mu$. Think of each column of $\bm{A}$ as a point in the $M$ dimensional space $ \mathbb{R}^M$. The minimum probability of error in distinguishing two points in $\mathbb{R}^M$, say $A_i$ and $A_j$ (that correspond to two columns of $\bm{A}$) scaled by $\mu$ is always bounded by $P_e ^{ i,j} \geq Q\left(\frac{\mu ||A_i-A_j||_2}{2}\right)$ where $||\cdot||_2$ denotes the $l_2$ norm and $Q(\cdot)$ is the Gaussian tail function defined as $Q(a)=\frac{1}{\sqrt{2\pi}}\int_{a}^{\infty}e^{-b^2/2}db $. For any non-adaptive strategy to attain a target probability of error $\epsilon$, we must have
\begin{equation}
 ||A_i-A_j||_2 \geq \frac{2Q^{-1}(\epsilon)}{\mu }\ \forall\ i,j\in \{1,2,\ldots,N\},\ i \neq  j
\end{equation}
Let $d=\frac{2Q^{-1}(\epsilon)}{\mu }$. 
Under $||A_i||_2 \leq \tau$ for some constant $\tau$, the problem translates to finding $N$ unique points in a sphere in $\mathbb{R}^M$ of radius $\tau$, that are each separated by at least $d$. As before, among all such matrices, (\textit{i.e.}, a collection of $N$ points in $\mathbb{R}^M$), we consider a matrix $\bm{A}$ with minimum $l_0$ cost and attempt to find a lower bound for it.

\newtheorem{corollary}{Corollary}
\begin{corollary}
Bisection-type adaptive strategies only require an $l_0$ cost of $\mathcal{O}(N)$ to reconstruct one--sparse signals when the number of measurements vary from $\Theta(\log_2(N))$ to $\Theta(N) $and are thus asymptotically dominated by non-adaptive strategies. 
\end{corollary}

In the next Section, we provide some preliminary lemmas and then we use them prove our theorems.




\section{Proofs of Theorems 1 and 2}


\subsection{Preliminary Results}
\newtheorem{lemma}{Lemma}
\begin{lemma}
 Let $r$ be an integer such that $1\leq r < M/2$. Also, let $p=\frac{r}{M}< 1/2$.  
\end{lemma}
\begin{equation}
\sum_{l=1}^r \binom{M}{l} \leq \frac{\binom{M}{r}}{1-2p}
\end{equation}
Moreover, if there exists a constant $\lambda$ such that $p\leq \lambda < 1/2$, then the partial summation is of the same order as the last term.
\begin{equation}\label{eq2}
\binom{M}{r} \leq \sum_{l=1}^r \binom{M}{l} \leq \frac{\binom{M}{r}}{1-2\lambda}
\end{equation}

\begin{IEEEproof}
First observe that 
\begin{equation}
  \frac{\binom{M}{r-1}}{\binom{M}{r}} = \frac{r}{M-r+1} < \frac{2r}{M}= 2p < 1
\end{equation}

Similarly, for $t=1,2,\cdots,r-2$,
\begin{equation}
    \frac{\binom{M}{r-t-1}}{\binom{M}{r-t}} = \frac{r-t}{M-r+t+1} < \frac{2r}{M}= 2p < 1
\end{equation}

Thus,
\begin{align}
\MoveEqLeft  \sum_{l=1}^r \binom{M}{l} \leq \binom{M}{r}\left(1+(2p)+(2p)^2+ \cdots+ (2p)^{(r-1)}  \right)  \nonumber \\
& \leq \binom{M}{r}\left(1+(2p)+(2p)^2+ \cdots \right) 
= \frac{\binom{M}{r}}{1-2p}
\end{align}

Note that, if $p\leq \lambda< 1/2$ ,(\ref{eq2}) follows.
\end{IEEEproof}


\begin{lemma}\cite[Lemma 4.7.1]{ash1990}
The binary entropy function is defined as $H(p)=-p\log_2(p)-(1-p)log_2(1-p)$  where $0 \leq p \leq 1$. Let $p=\frac{r}{M} \leq 1$. 

\begin{equation}
\frac{2^{M H(p)}}{\sqrt{8Mp(1-p)}} \leq \binom{M}{r} \leq \frac{2^{M H(p)}}{\sqrt{2\pi Mp(1-p)}}
\end{equation}
\end{lemma}


\begin{lemma}\cite[Lemma 2]{sahai2008price}
For an integer $d >1$, and $p<1/2$ 
\begin{equation}
H(p) \leq \frac{2d}{ln(2)} (p)^{1-\frac{1}{d}}
\end{equation}
\end{lemma}


\subsection{Binary Sensing Matrices}
 
\textit{Proof of Theorem 1:} Recall that we are required to bound $||\bm{A}||_{0,0}$ where the columns of $\bm{A}$ are chosen in ascending order of $l_0$ cost. Let $r_0$ be the integer such that
\begin{equation}
\sum_{l=1}^{r_0} \binom{M}{l} < N \leq \sum_{l=1}^{r_0+1} \binom{M}{l}
\end{equation}
For $l=1,2,\cdots,r_0$, there can be at most $\binom{M}{l}$ columns with $l$ ones, (\textit{i.e.}, $l_0$ cost $l$) and the remaining $\left(N-\sum_{l=1}^{r_{0}} \binom{M}{l}\right)$ columns have $l_0$ cost $(r_0+1)$.

If $M \leq \lfloor \log_2(N) \rfloor$, it is not possible to construct \textit{binary} matrices with $N$ unique columns. If $M=\lfloor \log_2(N)\rfloor+1$, then $r_0 \leq M/2$ and $||\bm{A}||_{0,0} \approx 0.5 N\log_2(N)$. As $M$ increases beyond $\lfloor \log_2(N)\rfloor+1$ for fixed $N$, $r_0$ decreases, and \emph{so does $||\bm{A}||_{0,0}$}. Thus, it is sufficient to prove the lower bound on $||\bm{A}||_{0,0}$ for  $M=T\log_2(N)$ where $T$ is a constant strictly greater than 1, as $||\bm{A}||_{0,0}$ for $T=1$ will only be higher. For $T>1$, we always have $r_0<M/2$ (required for \textit{Lemma 1}).  

Let $ N = c_M \sum_{l=1}^{r_0} \binom{M}{l}$ for some $c_M$. Let $p_0=\frac{r_0}{M}$. First, we show that if $M=T\log_2(N)$ ( $T>1$), then there exists a constant $\lambda$ such that $p_0\leq \lambda <1/2$. Note that,
\begin{equation}
 \frac{2^{M H(p_0)}}{\sqrt{8Mp_0(1-p_0)}} \overset{\textit{Lemma2}}{\leq}  \sum_{l=1}^{r_0} \binom{M}{l} < N
\end{equation}
Taking logarithm of both sides, 
\begin{equation}\label{eq11}
H(p_0)\leq \frac{\log_2(N)}{M} +\frac{\log_2(\sqrt{8r_0})}{M} < \frac{1}{T} + \frac{\log_2(\sqrt{8M})}{M}
\end{equation}
Note that, $\frac{\log_2(\sqrt{8M})}{M}$ is decreasing and tends to $0$ for large $M$. For any constant $\lambda'$ such that $(1/T)<\lambda'<1$, there always exists an $M=M(\lambda')$, such that
$\frac{\log_2(\sqrt{8M})}{M}<\lambda'-(1/T) $ for all $M >M(\lambda')$. Or, there exists a constant $\lambda=H^{-1}(\lambda')$ such that, $p_0 \leq \lambda < H^{-1}(1)=1/2$. This is required to apply \textit{Lemma 1}.

We now provide a lower bound on the $l_0$ cost of $\bm{A}$.
\begin{align}\label{eq3}
\MoveEqLeft ||\bm{A}||_{0,0} =\sum_{l=1}^{r_0} \binom{M}{l} \ l + \left(N-\sum_{l=1}^{r_0} \binom{M}{l}\right) \ (r_0+1)\nonumber\\
&\geq  r_0 \binom{M}{r_0} + \sum_{l=1}^{r_0} \binom{M}{l}(c_M-1)  (r_0+1)\nonumber\\
&\geq r_0 \ c_M \binom{M}{r_0} \overset{\textit{Lemma1}}{\geq} r_0\ c_M \sum_{l=1}^{r_0} \binom{M}{l}(1-2\lambda) \nonumber\\
&= (1-2\lambda) \ r_0\ N
\end{align}
Now, note that $r_0 < M$. Therefor either $r_0=\Theta(M)$ or $r_0 =\bm{o}(M)$. We show here that $r_0=\Theta(M)$. Look into the range of $c_M$.
\begin{align}\label{eq4}
\MoveEqLeft 1 < c_M \leq \frac{\sum_{l=1}^{r_0+1} \binom{M}{l}}{\sum_{l=1}^{r_0} \binom{M}{l}} <\frac{M+1}{r_0+1}<\frac{M}{2}
\end{align}
Using \textit{Lemma 1} and \textit{2},
\begin{equation}
N=c_M \sum_{l=1}^{r_0} \binom{M}{l}\leq \frac{c_M 2^{M H(p_0)}}{(1-2\lambda)\sqrt{\pi r_0}}
\end{equation}
Taking logarithm of both sides,
\begin{align} \label{eq12}
 \MoveEqLeft \frac{\log_2(N)}{M} \leq \frac{\log_2(c_M/((1-2\lambda)\sqrt{\pi r_0})}{M}+ H(p_0) \nonumber\\
 &\overset{(\ref{eq4})}{\leq}\frac{\log_2(M/(1-2\lambda))}{M}+ H(p_0)
\end{align}
Recall that $M=T\log_2(N)$ for some constant $T>1$. For large $M$, $\frac{\log_2(M/(1-2\lambda))}{M} \to 0$. For any positive constant $\eta' < (1/T)$, there always exist an $M=M(\eta')$ such that $\frac{\log_2(M/(1-2\lambda))}{M} <\eta' $ for all $M>M(\eta')$. \\Or, $H(p_0)\geq (1/T)-\eta' >0 $ for large $M$, which implies that, $r_0 \geq H^{-1}((1/T)-\eta') M=\eta M$ for some constant $\eta$.
Using this result in (\ref{eq3}), we finally obtain the lower bound,
\begin{equation}
||\bm{A}||_{0,0} \geq(1-2\lambda)\eta NM = C N \log_2(N)
\end{equation}
Here $C= (1-2\lambda)\eta T$ is a constant. Needless to say, the $l_0$ cost $||\bm{A}||_{0,0} \leq NM = T N\log_2(N)$ which is the total number of elements in an $M \times N$ matrix. Thus the sensing matrix is always dense,\textit{i.e.}, the $l_0$ cost is of same order as the total number of elements. Thus $||\bm{A}||_{0,0}=\Theta(N\log_2(N) )$\hfill $\IEEEQED$

\subsection{Real Valued Sensing Matrix in Noise}
\begin{figure}[!t]
\centering
\includegraphics[height=4in]{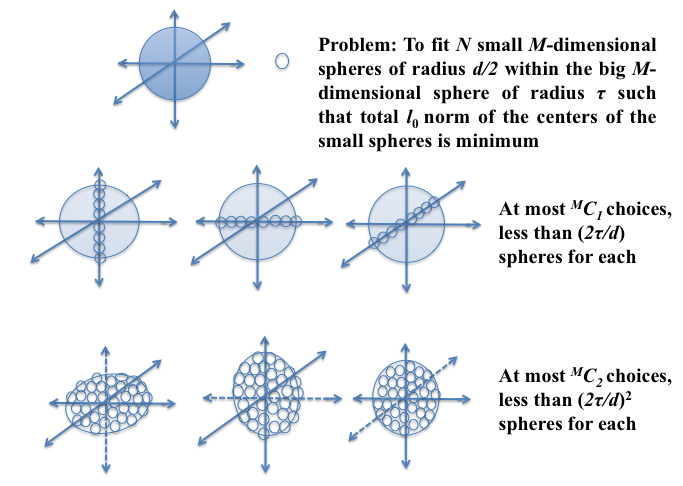}

\caption{A New Sphere Packing Problem - The figure illustrates the problem of finding $N$ unique columns of the sensing matrix $\bm{A}$ with minimum $l_0$ cost. The problem has an interesting interpretation as a sphere packing problem. For simplicity, we choose $M=3$ in this figure and then show how we can keep placing spheres of radius $d/2$ such that the total $l_0$ cost of the centres of the spheres (columns of $\bm{A}$) is minimum.}
\label{fig2}
\end{figure}
\textit{Proof of Theorem 2:}
The performance of non-adaptive reconstruction strategies relies on how well one can distinguish any two columns of sensing matrix $\bm{A}$, in the presence of noise. As already mentioned, we first consider the one--sparse case. We need to find $N$ unique points in a sphere in $\mathbb{R}^M$ dimensional space of radius $\tau$ that are each separated by at least $d =\frac{2 Q^{-1}(\epsilon)}{\mu}$. Assume we have $u_1$ points of $l_0$ cost $1$, $u_2$ points of $l_0$ cost $2$ and so on. Thus, $N=u_1+ u_2 +\cdots+u_{r_0'}$ and $||\bm{A}||_{0,0}= \sum_{l=1}^{r'}u_l \ l$.

Now consider the maximum possible number of points with $l_0$ cost equal to 1, and separated by distance $d$ in an $M$ dimensional sphere. Clearly there are $\binom{M}{1}$ possible choices of the non-zero dimension, and at most $\frac{2\tau}{d}$ points can be placed along each dimension. Thus $ u_1 \leq \binom{M}{1}\frac{2\tau}{d}$. 

Similarly, consider the maximum possible number of points with $l_0$ cost equal to $2$, and separated by distance $d$. There are $\binom{M}{2}$ possible choices of $2$ non-zero dimensions among $M$. For any such combination of $2$ dimensions, the problem is equivalent to fitting maximum number of $2D$ spheres of radius $d/2$ within a $2D$ sphere of radius  $\tau$. We would always have less than $\frac{\pi \tau^2}{\pi (d/2)^2}$ such points (or equivalently spheres) along any such combination of $2$ dimensions. Thus $ u_2 \leq \binom{M}{2}\left(\frac{2\tau}{d}\right)^2$.

For any $l \in \{1,2,\cdots,M\} $, the maximum number of points with $l_0$ cost equal to $l$ and separated by distance $d$ within the $M$ dimensional sphere of radius $\tau$ is always less than $\binom{M}{l}(\frac{2\tau}{d})^l $ as shown in Fig.\ref{fig2}. This follows since there are $\binom{M}{l}$ possible choices of the $l$ non-zero dimensions and for each such choice of $l$ non-zero dimensions, the number of $l$ dimensional spheres of radius $d/2$ that can be placed in a big $l$ dimensional sphere of radius $\tau$ is given by the ratio of the volumes of the two $l$ dimensional spheres. Thus $u_l \leq \binom{M}{l} (\frac{2\tau}{d})^l$. 

Choose $v_l=\binom{M}{l} \left(\frac{2\tau}{d}\right)^l \ \forall \ l \in {1,2 .. M}$. Let $r_0$ be the integer such that $\sum_{l=1}^{r_0} v_l < N \leq \sum_{l=1}^{r_0+1} v_l$.  Thus, for $\kappa >0$,\\
$ v_1+v_2+..v_{r_0}+\kappa= N =u_1+u_2+\cdots+u_{r_0'} $ 

First we prove that $r_0 < r_0'$. Assume $r_0 \geq r_0'$. Then,
\begin{align}
 \MoveEqLeft N=u_1+u_2+..+u_{r_0'} \leq v_1+v_2+..+v_{r_0'} \nonumber\\
 & \leq v_1+v_2..+v_{r_0} < N 
\end{align}
This is a contradiction. Thus, the $l_0$ cost can be lower bounded as shown below, using the fact that the summation puts more weights on the lower cost terms.
\begin{equation}
||\bm{A}||_{0,0}= \sum_{l=1}^{r_0'}u_l \ l \geq \sum_{l=1}^{r_0}v_l  l +\kappa (r_0+1)
\end{equation}
We follow a similar analysis to find a lower bound for $\sum_{l=1}^{r_0}v_l \ l +\kappa (r_0+1)$. Let $c_M\sum_{l=1}^{r_0} v_l = N$ and $p_0=r_0/M$. 

\begin{equation}
    H(p_0)+p_0\log_2\left(\frac{2\tau}{d}\right)\leq \frac{\log_2(N)}{M}+\frac{\log_2(\sqrt{8r_0})}{M}\to \frac{1}{T}
\end{equation}
Recall that, $\left(\frac{2\tau}{d}\right)$ is $\Theta(1)$. Thus, for large $M$, there exists a constant $\lambda<1/2$ such that, $p_0\leq \lambda < H^{-1}(1)=1/2  $.  We now provide a lower bound on the $l_0$ cost of $\bm{A}$.
\begin{align}\label{eq5}
\MoveEqLeft ||\bm{A}||_{0,0} \geq \sum_{l=1}^{r_0} v_l l + \sum_{l=1}^{r_0} v_l(c_M-1) \ (r_0+1)\nonumber\\
&\geq  r_0 \ c_M\ v_{r_0}  \overset{\textit{Lemma 1}}{\geq} r_0 N (1-2\lambda) 
\end{align}

Lastly, we show that $r_0=\Theta(M)$. As before, since it is already known that $r_0 <M$, either $r_0=\Theta(M)$ or $r_0=\bm{o}(M)$. Look into the range of $c_M$.

\begin{align}\label{eq6}
\MoveEqLeft 1 < c_M \leq \frac{\sum_{l=1}^{r_0+1} v_l}{\sum_{l=1}^{r_0} v_l}  <1+\frac{(M-r_0)(\frac{2\tau}{d})}{r_0+1}< M \frac{2\tau}{d}
\end{align}
Like the previous proof,
\begin{equation}
N =  c_M \sum_{l=1}^{r_0} \binom{M}{l}\left(\frac{2\tau}{d}\right)^l \leq \frac{c_M 2^{M H(p_0)}}{(1-2\lambda)\sqrt{\pi r_0}} \left(\frac{2\tau}{d}\right)^{r_0}
\end{equation}

Taking logarithm and using (\ref{eq6})
\begin{align}\label{eq7}
\MoveEqLeft \frac{\log_2(N)}{M}  \leq \frac{\log_2\left(\frac{2\tau M}{d(1-2\lambda)}\right)}{M} + H(p_0) + p_0\log_2\left(\frac{2\tau}{d}\right)
\end{align}
Now, if $p_0=r_0/M \to 0$ for large $M$, then all the terms in R.H.S. of (\ref{eq7}) would tend to $0$.  But since $M=T\log_2(N)$ for some constant $T \geq 1$, the L.H.S. is always a non-zero constant $1/T$. Thus we arrive at a contradiction. Thus, we have $r_0 \geq \eta M$ for some constant $\eta=\Theta(1)$.
Using  this result in (\ref{eq5}), we obtain,
\begin{equation}
    ||\bm{A}||_{0,0} \geq (1-2\lambda)\eta N M\geq C N \log_2(N)
\end{equation}
As $M=\Theta(\log_2(N))$, we have $||\bm{A}||_{0,0} =\Theta(N\log_2(N))$. \hfill $\IEEEQED$
%
\subsection{Comparison with Adaptive Sensing}
\textit{Proof of Corollary: }
Consider the $l_0$ cost of the bisection-type adaptive sensing strategies as described in \cite{davenport2012compressive}. As in Fig.\ref{fig1}, first the total length $\{1,2,\cdots,N\}$ is divided into two equal blocks, and any one measurement is acquired. Based on the received measurement value, the active block (set of possible non-zero locations) is narrowed down. That blocks is again divided into two parts and any one part is sensed. For such a sensing strategy,
\begin{equation}
||\bm{A}||_{0,0} \leq \left(\frac{N}{2} + \frac{N}{4} + \cdots \right) \leq N
\end{equation} 
Thus, adaptive sensing matrices require an $l_0$ cost of at most $\mathcal{O}(N)$ for signal reconstruction and are thus sparse. Clearly, this upper bound is asymptotically dominated by $N\log_2(N)$. 

Note that, since we have already allowed the values of the non-zero elements of $\bm{x}$ to be sufficiently high, it is guaranteed (from \cite{malloy2014near}) to succeed in the presence of noise.

\subsection{Discussion on higher orders of M}
We consider the case of reconstruction of one--sparse signals using \textit{binary} sensing matrices, when $M$ is between $\log_2(N)$ and $N$. From (\ref{eq11}), we note that $p_0 \to 0$ if $\log_2(N)/M \to 0$ for large $M$ and $N$. We can always find a constant $\lambda$ arbitrarily small, such that $p_0\leq \lambda <1/2$. The binomial summation in \textit{Lemma 1} is practically equal to the largest term. From (\ref{eq3})
\begin{equation}
  ||\bm{A}||_{0,0} \geq c_M r_0 \binom{M}{r_0} \geq r_0 N(1-2\lambda) \approx r_0 N
\end{equation}
In general, note that when $M=\Theta(\log_2(N))$, then $r_0=\Theta(M)$ and $||\bm{A}||_{0,0}=\Theta(N\log_2(N))$. As $M$ increases, the lower bound on $l_0$ cost decreases. When $M$ is of higher order, say $M=N$, then $r_0=1$ and we have a reduced $l_0$ cost $N$ using simply an identity sensing matrix. From (\ref{eq12}),
\begin{align}
 \frac{\log_2(N/c_M)}{M}\leq H(r_0/M)\overset{\textit{Lemma 3}}{\leq} \frac{2d}{ln(2)} \left(\frac{r_0}{M}\right)^{1-\frac{1}{d}}\\
 \left(\frac{ln2 \ \log_2(N/c_M)}{2d}\right)^{d/(d-1)} M^{-1/(d-1)}\leq r_0\ \ \ \ \ \
\end{align}
This leads to the following bound.
\begin{equation}
||\bm{A}||_{0,0}
\geq C \max_{d} \frac{N}{ M^{1/(d-1)}} \left(\frac{ln2 \log_2(N/c_M)}{2d}\right)^{d/d-1} 
\end{equation}
Similar bounds may be derived for real-valued matrices.


\section{Conclusions}
This paper clearly shows an advantage of adaptivity in terms of $l_0$ cost of sensing. It establishes a lower bound on the $l_0$ cost that holds for any non-adaptive strategy, whether it is random or not.  $K$--sparse signals, we can argue that if a matrix does not have unique columns and fails to reconstruct one--sparse signals, it naturally fails to reconstruct a $K$--sparse signal. If a matrix has two similar columns say $\{i_1\}$ and $\{i_2\}$, then two exactly $K$ sized supports differing only at $\{i_1\}$ and $\{i_2\}$ become indistinguishable. Thus, the lower bound on the $l_0$ cost of $\bm{A}$  cannot be less than that in the case of one--sparse signals. Thus, $||\bm{A}||_{0,0}=\Theta\left(N K\log_2(N)\right)$ when $M=\Theta(K\log_2(N))$. For small values of $K$,  this still asymptotically dominates $ \mathcal{O}(KN)$ which is the $l_0$ cost for bisection-type adaptive strategies.  
\section*{Acknowledgment}
This work was supported in part by Systems on Nanoscale Information fabriCs (SONIC), one of the six SRC STARnet Centers, sponsored by MARCO and DARPA. We also acknowledge the support of NSF-CCF-1350314 (NSF Career Award). S Dutta was also supported by a Dean's Fellowship from Carnegie Mellon University.

We also thank Praveen Venkatesh, Haewon Jeong, Yaoqing Yang and Aarti Singh for their suggestions on the paper.


\ifCLASSOPTIONcaptionsoff
  \newpage
\fi



\bibliographystyle{IEEEtran}
\bibliography{IEEEabrv,sample}
\end{document}